\documentclass[aps,preprint,showpacs,floats,epsf,epsfig,nofootinbib,12pt]{revtex4}
\usepackage{graphicx}% Include figure files
\usepackage{epsfig}
\usepackage[dvips]{color}

\def\beq{\begin{eqnarray}}
\def\eeq{\end{eqnarray}}
\def\non{\nonumber}

\def\la{\langle}
\def\ra{\rangle}

\begin{document}

\title{ Study on  the  weak decay between two heavy baryons $ \mathcal{B}_i(\frac{1}{2}^+)\to  \mathcal{B}_f(\frac{3}{2}^+)$ in the light-front quark model}

\vspace{1cm}

\author{ Fang Lu$^{1}$, Hong-Wei Ke$^{1}$   \footnote{khw020056@tju.edu.cn, corresponding author}, Xiao-Hai
Liu$^1$\footnote{xiaohai.liu@tju.edu.cn} and Yan-Liang Shi$^2$\footnote{ys6085@princeton.edu}}

\affiliation{  $^{1}$ School of Science, Tianjin University, Tianjin 300072, China\\
  $^{2}$ Princeton Neuroscience Institute, Princeton University, Princeton, NJ 08544, USA}

\vspace{12cm}

\begin{abstract}
In this work, we study the weak decay between two heavy baryons $ \mathcal{B}_i(\frac{1}{2}^+)\to  \mathcal{B}_f(\frac{3}{2}^+)$ in the light-front quark model where three-quark picture is employed for baryon.
We derive general form of transition amplitude of $ \mathcal{B}_i(\frac{1}{2}^+)\to  \mathcal{B}_f(\frac{3}{2}^+)$, and  analyze two specific cases of transitions:  the weak decays of single heavy baryon $\Sigma_{b} \to \Sigma_{c}^*$ and the decays of double-charmed  baryon $\Xi_{cc}\to \Sigma_{c}^*(\Xi_{c}^*)$.
We compute the hadronic form factors for the transitions
and apply them to study the decay widths of the semi-leptonic $\mathcal B_i(\frac{1}{2}^+)\to\mathcal B_f(\frac{3}{2}^+) l\bar{\nu}_l$
and non-leptonic $\mathcal B_i(\frac{1}{2}^+)\to\mathcal B_f(\frac{3}{2}^+)M$.
Previously we studied the transition $\Sigma_{b} \to \Sigma_{c}^*$  with the quark-diquark picture of baryon  in the light-front quark model. Here we revisit this transition with three-quark picture of baryon.
At the quark level, the transition $\Sigma_{b} \to \Sigma_{c}^*$ is induced by the $b\rightarrow c$ transition.The subsystem of the two unchanged light quarks which possesses definite and same spin in initial and final state can be viewed as a spectator, so the spectator approximation can be applied directly.  For the weak decay of doubly charmed baryon $\Xi_{cc}$, a $c$ quark decays to a light quark $q_1$, so both the initial state $cc$ and final state $q_1q_2$ ($q_1$ and the original $q_2$ in initial state may be the same flavor quarks) which possess definite spin are no longer spectators. A rearrangement of quarks for initial and final states is adopted to isolate the unchanged subsystem $cq_2$  which can be viewed as the spectator approximately.
Future measurements on these channels will constrain the nonperturbative parameter in the wavefunctions  and test the model predictions.


\end{abstract}

\maketitle

\section{Introduction}
Over the last decade,  a great interest has been aroused in the field of hardon physics,  especially heavy baryons. Significant progresses have been achieved in both experiment and theory. For example, the LHCb collaboration observed the doubly charmed baryon $\Xi_{cc}^{++}$ in the final state $\Lambda_cK^{-}\pi^+\pi^+$ \cite{LHCb:2017iph} and it has been confirmed in the
decays $\Xi_{cc}^{++}\to\Xi_{c}^+\pi^+$and $\Xi_{cc}^{++}\to\Xi_{c}^{'+}\pi^+$ \cite{LHCb:2018pcs,LHCb:2022rpd}. On the theory side, multiple approaches have been developed to study the physical properties of heavy baryons, such as decay width. For example, in Refs. \cite{Yu:2017zst,Wang:2017mqp,Zhao:2018zcb} the authors employed light-front quark model (LFQM)
to explore the weak decays of doubly heavy baryons. In Ref. \cite{Hu:2019bqj} the weak decays of doubly heavy baryons were studyed within light-cone sum rules. In retrospect, weak decays of the heavy baryons were also studied under the heavy quark limit \cite{Korner:1992uw}, the relativistic quark model with quark-diquark picture \cite{Ebert:2006rp}, the
relativistic three-quark model \cite{Ivanov:1996fj} and the Bethe-Salpeter approach \cite{Ivanov:1998ya}.

  In Refs. \cite{Yu:2017zst,Wang:2017mqp,Zhao:2018zcb,Ke:2007tg,Wei:2009np,Ke:2012wa,Ke:2017eqo}
the LFQM was extended to study the weak decays of the heavy baryons with the
quark-diquark picture. However,  although the quark-diquark picture is an effective approximation when the diquark is a spectator in the transition, it is not very suitable for studying the decay where the diquark will be broken. Instead, three-quark picture was employed to study the weak decays between two heavy baryons with $J^P=\frac{1}{2}^+$   in Refs. \cite{Ke:2019smy,Ke:2019lcf,Ke:2021pxk,Li:2021kfb}. In Ref. \cite{Ke:2017eqo} the transition between $J^P=\frac{1}{2}^+$ heavy baryon ($\mathcal{B}_i(\frac{1}{2}^+)$) and $J^P=\frac{3}{2}^+$ one ($\mathcal{B}_f(\frac{3}{2}^+)$) was studied with the
quark-diquark picture. In this work we  employ the three-quark picture to study the transition $\mathcal{B}_i(\frac{1}{2}^+)\to \mathcal{B}_f(\frac{3}{2}^+)$ within the LFQM framework. The
LFQM  is a relativistic quark model which
has been applied to study transitions among mesons
\cite{Jaus,Ji:1992yf,Cheng:1996if,Cheng:2003sm,Hwang:2006cua,Lu:2007sg,Choi:2007se,
Li:2010bb,Ke:2009ed,Wei:2009nc,Ke:2009mn,Ke:2011fj,Ke:2011mu,Ke:2011jf} and has been extended to the case of baryon decay \cite{Ke:2017eqo,Ke:2007tg,Wei:2009np,Ke:2012wa,Wang:2017mqp,Zhao:2018zcb,Yu:2017zst}.
We derive general forms of transition amplitude  $\mathcal{B}_i(\frac{1}{2}^+)\to \mathcal{B}_f(\frac{3}{2}^+)$. Then we revisit the transition $\Sigma_{b}\to\Sigma_{c}^*$ and study the relative weak decays of the double charmed baryon $\Xi_{cc}$.

For the weak decay of single heavy baryon, the $b$ quark in the initial state would
transit into the $c$ quark in the final state by emitting $W$ bosons.
The two light quarks do not take part in the process of the weak decay and the subsystem of the two light quarks can be regarded as a spectators.
Under the three-quark picture, the three quarks are regarded as independent individuals.
Although the two light quarks are no longer point-like diquark, the subsystem where
they reside still has a definite spin, color, isospin and all the quantum numbers of the subsystem keep unchanged so it
is treated as the spectator.
For the weak decay of doubly charmed baryon ($ccq_2$), one  $c$ quark in the initial state
decays to a lighter quark $q_1$ by emitting $W$ bosons via the weak interaction. The resulting
new quark forms a subsystem ($q_1q_2$) with the light quark $q_2$ which comes from the initial state. As a result,  both
$cc$ pair in the initial state and  $q_1q_2$ in the final state which have  definite spin cannot be treated as the spectators in the decay. A rearrangement of quarks for the initial and final baryons is necessary to isolate the unchanged subsystem $cq_2$, which can be viewed as the spectator approximately.

In LFQM, we need to specify the vertex functions of the initial state and the final state according to  the total spin and parity of the baryon. Here the quantum numbers of the initial state and the final state are ${1\over 2}^+$ and ${3\over 2}^+$, respectively.  We adopt  the vertex function of ${1\over 2}^+$ baryon in Ref. \cite{Ke:2019smy} and construct the three-body vertex function of ${3\over 2}^+$ in analog to Ref. \cite{Ke:2019smy}. Then we derive the transition matrix element, extract the form factors and compute them numerically.

Using these form factors we calculate the associated non-leptonic decays and semileptonic decays. The
 leptons are not involved in the strong interaction which means the semileptonic decay is
 less contaminated by the non-perturbative QCD effect, hence the study on semileptonic
 decay can be very helpful to constrain the model parameters and test model predictions. For the non-leptonic two-body decays ${\Sigma_b}\to\Sigma_c^*M$, ${\Xi_{cc}}\to\Sigma_c^*M$ and ${\Xi_{cc}}\to\Xi_c^{*}M$,
by neglecting  the interaction between the final states,  the transition element is factorized to a product of the meson decay constant and the transition amplitude  $\mathcal{B}_i(\frac{1}{2}^+)\to \mathcal{B}_f(\frac{3}{2}^+)$ under the factorization assumption.
Future measurement on these channels will be necessary to further test model predictions and elucidate the underlying mechanism of heavy hadron decay.

This paper is organized as follows:  in
section II we present the vertex functions of the heavy flavor baryons,
and  derive the form factors of the transition $\mathcal B_i(\frac{1}{2}^+)\to\mathcal B_f(\frac{3}{2}^+)$  in
the LFQM. In section III we
present numerical results for the transition $\mathcal B_i(\frac{1}{2}^+)\to\mathcal B_f(\frac{3}{2}^+)$ along with all
necessary input parameters.  We calculate the form factors and the decay widths of related semi-leptonic and non-leptonic decays.
Section IV is devoted to our conclusions and discussions.

\section{ $\mathcal{B}_i(\frac{1}{2}^+)\to\mathcal B_f(\frac{3}{2}^+)$ in the light-front quark model}

In this paper we study the transition $\mathcal{B}_i(\frac{1}{2}^+)\to\mathcal B_f(\frac{3}{2}^+)$ with the three-quark picture of baryon in LFQM. We focus on the weak decay of single heavy baryon $\Sigma_{b}$ to $\Sigma_{c}^*$ and the doubly charmed baryon $\Xi_{cc}$ to single charmed baryon $\Sigma_{c}^*$ or $\Xi_{c}^*$.

\subsection{the vertex functions}

In Ref. \cite{Ke:2019smy}, the vertex function of a baryon
$\mathcal B$ with the total spin $S=1/2$ and total momentum $P$  was
discussed\footnote{Since the orbital angle momentum we study is 0, the total angle momentum $J$ is equal to the total spin $S$ of the baryon. }. For a single heavy baryon it could be expressed as
\begin{eqnarray}\label{baryonv1}
  && |\mathcal B(P,S,S_z)\rangle=\int\{d^3\tilde p_1\}\{d^3\tilde
p_2\}\{d^3\tilde p_3\} \,
  2(2\pi)^3\delta^3(\tilde{P}-\tilde{p_1}-\tilde{p_2}-\tilde{p_3}) \nonumber\\
 &&\times\sum_{\lambda_1,\lambda_2,\lambda_3}\Psi_{S_{q_1q_2}}^{SS_z}(\tilde{p}_1,\tilde{p}_2,\tilde{p}_3,\lambda_1,\lambda_2,\lambda_3)
  \mathcal{C}^{\alpha\beta\gamma}\mathcal{F}_{Q,q_1,q_2} \left|\right. Q_{\alpha}(p_1,\lambda_1)q_{1\beta}(p_2,\lambda_2)q_{2\gamma}(p_3,\lambda_3)\rangle
  ,\end{eqnarray}
where $Q$ denotes heavy quark ($b$ or $c$), $q_{1}(q_{2})$ denotes the flavor of the light quark,  $\lambda_i$ and $p_i\, (i=1,2,3)$ are helicities and
light-front momenta of the quarks,
$\mathcal{C}^{\alpha\beta\gamma}$ and $\mathcal{F}_{Q,q_1,q_2}$
are the color and flavor factors. The total spin of $q_1$ and
$q_2$ is denoted to $S_{q_1q_2}$ which is 0 for $\Lambda_b\,
( \Lambda_c)$ and 1 for $\Sigma_b\, ( \Sigma_c)$, respectively.
In light-front approach, the on-mass-shell light-front momentum $p$ is defined as
\begin{equation}
\tilde{p}=(p^{+},p_{\perp}),  \quad, p_{\perp} =(p^{1},p^{2}), \quad p^{-} =\frac{m^2+p^2_{\perp}}{p^{+}} \ , \quad
\{d^3p\} = \frac{dp^{+}d^2p_{\perp}}{2(2\pi)^3} \ .
\end{equation}
The momentum-space wave function $\Psi^{SS_z}_0$ and $\Psi^{SS_z}_1$\cite{Ke:2019smy,Tawfiq:1998nk} are
\begin{eqnarray}\label{wave_1}
\Psi^{SS_z}_0(\tilde{p}_i,\lambda_i)=&&A_0 \bar
u(p_3,\lambda_3)[(\bar
P\!\!\!\!\slash+M_0)\gamma_5]v(p_2,\lambda_2)\bar
u(p_1,\lambda_1)  u(\bar P,S) \phi_\mathcal{B}(x_i,k_{i\perp}),\nonumber\\
\Psi^{SS_z}_1(\tilde{p}_i,\lambda_i)=&&A_1 \bar
u(p_3,\lambda_3)[(\bar
P\!\!\!\!\slash+M_0) \gamma_{\perp}^\beta]v(p_2,\lambda_2)\bar
u(p_1,\lambda_1) \gamma_{\perp\beta}\gamma_{5} u(\bar P,S) \phi_\mathcal{B}(x_i,k_{i\perp}).
\end{eqnarray}

For a single heavy baryon $\mathcal B$ with total spin $S=3/2$
the vertex function is the same as that in Eq. (\ref{baryonv1}) except $\Psi^{SS_z}_1$ is replaced by $\Psi^{'SS_z}_1$ \cite{Tawfiq:1998uq}, which is given by
\begin{eqnarray}\label{wave_2}
\Psi^{'SS_z}_1(\tilde{p}_i,\lambda_i)=&&A'_1 \bar
u(p_3,\lambda_3)[(\bar
P\!\!\!\!\slash+M_0) \gamma_{\perp}^\alpha]v(p_2,\lambda_2)\bar
u(p_1,\lambda_1) u_{\alpha}(\bar
P,S)\phi_\mathcal{B}(x_i,k_{i\perp}).
\end{eqnarray}

With the normalization of the state  $\mathcal B$ \beq\label{A121}
 \la
 \mathcal B (P',S',S'_z)|\mathcal B (P,S,S_z)\ra=2(2\pi)^3P^+
  \delta^3(\tilde{P}'-\tilde{P})\delta_{S'S}\delta_{S'_zS_z},
 \eeq
and \beq\label{A122}
\int(\prod^3_{i=1}\frac{dx_id^2k_{i\perp}}{2(2\pi)^3})2(2\pi)^3\delta(1-\sum
x_i)\delta^2(\sum
k_{i\perp})\phi^*_\mathcal{B}(x_i,k_{i\perp})\phi_\mathcal{B}(x_i,k_{i\perp})=1,
 \eeq
we can compute the factors $A_0$, $A_1$ and $A'_1$:
\begin{eqnarray}
A_0
&&=\frac{1}{4\sqrt{P^+M_0^3(m_1+e_1)(m_2+e_2)(m_3+e_3)}},\nonumber\\A_1
&&=\frac{1}{4\sqrt{3P^+M_0^3(m_1+e_1)(m_2+e_2)(m_3+e_3)}},\nonumber\\
A'_1
%&&=\frac{1}{4\sqrt{2P^+(M_0m_1+p_1\cdot\bar{P})(M_0m_2+p_2\cdot\bar{P})(M_0m_3+p_3\cdot\bar{P})}}\nonumber\\
&&=\frac{1}{4\sqrt{2P^+M_0^3(m_1+e_1)(m_2+e_2)(m_3+e_3)}},
\end{eqnarray}
where $p_i\cdot\bar{P}=e_iM_0\,(i=1,2,3)$ is used, $\bar{P}=p_1+p_2+p_3$  and $e_i$ is defined in Eq. (\ref{e_i}).

For the weak decay of doubly charmed baryon, the vertex functions
with total spin $S=1/2$ and momentum $P$ is
\begin{eqnarray}\label{eq:lfbaryon}
  && |\mathcal B(P,S,S_z)\rangle=\int\{d^3\tilde p_1\}\{d^3\tilde
p_2\}\{d^3\tilde p_3\} \,
  2(2\pi)^3\delta^3(\tilde{P}-\tilde{p_1}-\tilde{p_2}-\tilde{p_3}) \nonumber\\
 &&\times\sum_{\lambda_1,\lambda_2,\lambda_3}\Psi_{S_{QQ}}^{SS_z}(\tilde{p}_1,\tilde{p}_2,\tilde{p}_3,\lambda_1,\lambda_2,\lambda_3)
  \mathcal{C}^{\alpha\beta\gamma}\mathcal{F}_{Q,Q,q_2} \left|\right. Q_{\alpha}(p_1,\lambda_1)Q_{\beta}(p_2,\lambda_2)q_{2\gamma}(p_3,\lambda_3)\rangle.
  \end{eqnarray}

In order to describe the momenta of the constituent quarks, the intrinsic variables $(x_i, k_{i\perp})$ (
$i=1,2,3$) are introduced through
\begin{eqnarray}
&&p^+_i=x_i P^+, \qquad p_{i\perp}=x_i P_{\perp}+k_{i\perp}
 \qquad x_1+x_2+x_3=1, \qquad k_{1\perp}+k_{2\perp}+k_{3\perp}=0,
\end{eqnarray}
{where $x_i$ is the momentum fraction with the constraint
$0<x_1, x_2, x_3<1$. The variables $(x_i, k_{i\perp})$ are
Lorentz-invariant  since they are independent of the total momentum of the hadron. The invariant mass square $M_0^2$ is
defined as} a function of the internal variables $x_i$ and $k_{i\perp}$:
 \begin{eqnarray} \label{eq:Mpz}
  M_0^2=\frac{k_{1\perp}^2+m_1^2}{x_1}+
        \frac{k_{2\perp}^2+m_2^2}{x_2}+\frac{k_{3\perp}^2+m_3^2}{x_3}.
 \end{eqnarray}

The internal momenta are defined  as
 \beq
 k_i=(k_i^-,k_i^+,k_{i\bot})=(e_i-k_{iz},e_i+k_{iz},k_{i\bot})=
  (\frac{m_i^2+k_{i\bot}^2}{x_iM_0},x_iM_0,k_{i\bot}).
 \eeq

It is easy to obtain
 \begin{eqnarray}
  e_i&=&\frac{x_iM_0}{2}+\frac{m_i^2+k_{i\perp}^2}{2x_iM_0}
 ,\non\\
 k_{iz}&=&\frac{x_iM_0}{2}-\frac{m_i^2+k_{i\perp}^2}{2x_iM_0},
 \label{e_i}
 \end{eqnarray}
{where $e_i$ is the energy of the $i$-th constituent, and
they obey the condition $e_1+e_2+e_3=M_0$. The transverse $k_{i\bot}$ and $z$ direction $k_{iz}$ components  constitute a momentum vector $\vec k_i=(k_{i\bot}, k_{iz})$.}

The spatial wave function  $\phi_\mathcal{B}(x_i, k_{i\perp})$  \cite{pentaquark1,pentaquark2} is defined as
 \beq\label{A122}
\phi_\mathcal{B}(x_1,x_2,x_3,k_{1\perp},k_{2\perp},k_{3\perp})=\frac{e_1e_2e_3}{x_1x_2x_3M_0}
\varphi(\overrightarrow{k}_1,\beta_1)\varphi(\frac{\overrightarrow{k}_2-\overrightarrow{k}_3}{2},\beta_{23})
 \eeq
with
$\varphi(\overrightarrow{k},\beta)=4(\frac{\pi}{\beta^2})^{3/4}{\rm
exp}(\frac{-k_z^2-k^2_\perp}{2\beta^2})$, where $\beta$ is a non-perturbative parameter that characterizes the shape of wave function. We will discuss the parameter selection in section \ref{section_numerical}.

\subsection{the  form factors of $\mathcal B_i(\frac{1}{2}^+)\to\mathcal B_f(\frac{3}{2}^+)$ in LFQM}

{The form factors for the transition from initial heavy baryon $\mathcal B_i(\frac{1}{2}^+)$  (i.e. $|\mathcal B(P,1/2,S_z)$)  to final heavy baryon $\mathcal B_f(\frac{3}{2}^+)$ ($|\mathcal B(P',3/2,S'_z)\rangle$) are defined as
\begin{eqnarray}\label{s1}
&& \langle\mathcal B_f(\frac{3}{2}^+) \mid
\bar{Q_2}
\gamma^{\mu} (1-\gamma_{5}) Q_1 \mid \mathcal B_i(\frac{1}{2}^+)\ra
 = \non \\&& \bar{u}_{\alpha}(P',S'_z) \left[ \gamma^{\mu} P^\alpha \frac{f_{1}(q^{2})}{M_{\mathcal B_i}}
 + \frac{f_{2}(q^{2})}{M^2_{\mathcal B_i}} P^\alpha P^\mu+
 \frac{f_{3}(q^{2})}{M_{\mathcal B_i}M_{\mathcal B_f}} P^\alpha P'^\mu+f_4(q^{2})g^{\alpha \mu}
 \right] u(P,S_z)- \nonumber \\
 &&\bar u_{\alpha}(P',S'_z)\left[\gamma^{\mu} P^\alpha \frac{g_{1}(q^{2})}{M_{\mathcal B_i}}
 + \frac{g_{2}(q^{2})}{M^2_{\mathcal B_i}} P^\alpha P^\mu+
 \frac{g_{3}(q^{2})}{M_{\mathcal B_i}M_{\mathcal B_f}} P^\alpha P'^\mu+g_4(q^{2})g^{\alpha \mu}
 \right]\gamma_{5} u(P,S_z),
\end{eqnarray}
where  momentum $q \equiv P-P'$. $Q_1$ and $Q_2$ denote heavy quark operators (See Fig. \ref{t1}).  $M_{\mathcal B_i}$ and $M_{\mathcal B_f}$ represent the masses of heavy baryon $\mathcal B_i$ and $\mathcal B_f$, respectively. For the deday of doubly charmed baryon  $Q_2$ should be replaced by $q_1$ (See Fig. \ref{t2}).

Here the momenta of initial and final baryons $P$ and $P'$ obey the on-shell relations $E^{(')2}=P^{(')2}+M^{(')2}$. However, in Eq. (\ref{wave_1}) and (\ref{wave_2}) the spinors are function of  {$\bar P$ and $\bar P'$, which  are the sums of the momenta of the involved constituent quarks and do not obey physical on-shell relations. To reconcile the conflict, here we follow the approach in the previous study \cite{Ke:2017eqo} and assume form factors ( $f_1$, $f_2$, $f_3$, $f_4$, $g_1$, $g_2$,	$g_3$, $g_4$) are the same in both physical and unphysical form. Then Eq. (\ref{s1}) is re-written as the following equation  where the spinors  are off-shell:
%
% define and explain \bar P, \bar P', why they are un-physical.
% why we can assume form factor are invariant ?
%
\begin{eqnarray}\label{s2}
&& \langle\mathcal B_f(\frac{3}{2}^+) \mid
\bar{Q_2}
\gamma^{\mu} (1-\gamma_{5}) Q_1 \mid \mathcal B_i(\frac{1}{2}^+) \ra
= \non \\&&\bar{u}_{\alpha}(\bar P',S'_z) \left[ \gamma^{\mu}\bar P^\alpha \frac{f_{1}(q^{2})}{M_{\mathcal B_i}}
 + \frac{f_{2}(q^{2})}{M^2_{\mathcal B_i}}\bar P^\alpha \bar P^\mu+\frac{f_{3}(q^{2})}{M_{\mathcal B_i}M_{\mathcal B_f}}\bar P^\alpha \bar P'^\mu
 +f_4(q^{2})g^{\alpha \mu}
 \right] u(\bar P,S_z) -\nonumber \\
 &&\bar u_{\alpha}(\bar P',S'_z)\left[\gamma^{\mu}\bar P^\alpha \frac{g_{1}(q^{2})}{M_{\mathcal B_i}}
 + \frac{g_{2}(q^{2})}{M^2_{\mathcal B_i}}\bar P^\alpha \bar P^\mu+\frac{g_{3}(q^{2})}
 {M_{\mathcal B_i}M_{\mathcal B_f}}\bar P^\alpha \bar P'^\mu+g_4(q^{2})g^{\alpha \mu}
 \right]\gamma_{5} u(\bar P,S_z).
\end{eqnarray}

\begin{center}
\begin{figure}[htb]
\begin{tabular}{cc}
\scalebox{0.7}{\includegraphics{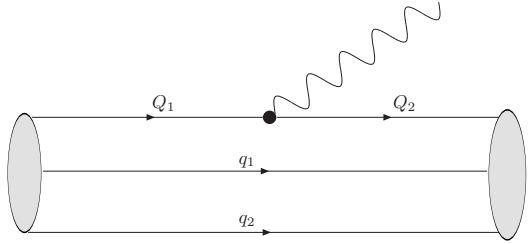}}
\end{tabular}
\caption{ The Feynman diagram for
 single heavy baryon weak decay, where $\bullet$ denotes
$V-A$ current vertex.}\label{t1}
\end{figure}
\end{center}

\subsubsection{the transition $\Sigma_b\to\Sigma^*_c$}

{The lowest order Feynman diagram responsible for
the $\Sigma_b\to\Sigma^*_c$ ( $ \Sigma_b$ and $\Sigma^*_c$ are $\mathcal B_i(\frac{1}{2}^+)$ and $\mathcal
B_f(\frac{3}{2}^+)$, respectively) weak decay is shown in
Fig. \ref{t1}. Following the approach given in
Refs. \cite{pentaquark1,pentaquark2,Ke:2007tg,Ke:2012wa} the
transition matrix element can be calculated with the vertex
functions of $\mid \mathcal B_i(\frac{1}{2}^+) \ra$ and $\mid
\mathcal B_f(\frac{3}{2}^+)\ra$},

\begin{eqnarray}\label{s3}
&& \langle\mathcal B_f(\frac{3}{2}^+) \mid
\bar{Q_2}
\gamma^{\mu} (1-\gamma_{5}) Q_1 \mid \mathcal B_i(\frac{1}{2}^+)  \ra = \nonumber \\
 && \int\frac{\{d^3 \tilde p_2\}\{d^3 \tilde p_3\}\phi_{B_f}^*(x',k'_{\perp})
  \phi_{B_i}(x,k_{\perp}){\rm Tr}[\gamma_{\perp}^\alpha(\bar{P'}\!\!\!\!\!\slash+M_0')(p_3\!\!\!\!\!\slash+m_3)
  (\bar{P}\!\!\!\!\!\slash+M_0)\gamma_{\perp}^\beta(p_2\!\!\!\!\!\slash-m_2)]}{16\sqrt{6p^+_1p'^+_1{P}^+{P'}^+M_0^3M_0'^3(m_1+e_1)
 (m_2+e_2)(m_3+e_3)(m_1'+e_1')
 (m_2'+e_2')(m_3'+e_3')}}\nonumber \\
  &&\times  \bar{u}_{\alpha}(\bar{P'},S'_z)
  (p_1\!\!\!\!\!\slash'+m'_1)\gamma^{\mu}(1-\gamma_{5})
  (p_1\!\!\!\!\!\slash+m_1)\gamma_{\perp\beta}\gamma_{5}  u(\bar{P},S_z),
\end{eqnarray}
where for the weak decay of single heavy baryons,
 \beq
m_1=m_b, \qquad m'_1=m_c, \qquad m_2=m_{q_1}, \qquad m_3=m_{q_2}.
 \nonumber\eeq
$p_1$ denotes the four-momentum of the heavy
quark $b$, $p'_1$ denotes the four-momentum of the quark $c$,
 $P$ ($P'$) stands as the four-momentum of $\mathcal
B_i$ ($\mathcal B_f$). Setting
$\tilde{p}_2=\tilde{p}'_2$, $\tilde{p}_3=\tilde{p}'_3$ we have
 \beq
 x'_{i}=\frac{P^+}{P^{'+}}x_{i}, \qquad  k'_{1\perp}=k_{1\perp}-(1-x_1)q_{\perp}, \qquad k'_{2\perp}=k_{2\perp}+x_2q_{\perp}, \qquad k'_{3\perp}=k_{3\perp}+x_3q_{\perp}.
 \eeq

Multiplying the following expressions $\bar u(\bar
P,S_z)\gamma_{\mu}\bar P^\xi {u}_{\xi}(\bar P',S'_z)$, $\bar
u(\bar P,S_z)\bar P'_{\mu}\bar P^\xi{u}_{\xi}(\bar P',S'_z)$,
$\bar u(\bar P,S_z) \bar P_{\mu}\bar P^\xi{u}_{\xi}(\bar
P',S'_z)$, $\bar u(\bar P,S_z) g_{\mu}^{\xi} {u}_{\xi}(\bar
P',S'_z)$ to the right sides of both
Eq. (\ref{s2}) and Eq. (\ref{s3}) and then summing over
the polarizations of all states, we can obtain four algebraic equations and  each equation contains the form factors $f_{1}$, $f_{2}$, $f_{3}$ and $f_{4}$. Solving these equations, we can get the explicit expressions of these form factors  $f_{i}$($i=1,2,3,4$) (See Appendix $A$ for detail).

Similarly, multiplying these expressions  $ \bar u(\bar P,S_z)\gamma_{\mu}\bar P^\xi
\gamma_{5}{u}_{\xi}(\bar P',S'_z)$, $\bar u(\bar P,S_z)\bar
P'_{\mu}\bar P^\xi\gamma_{5}{u}_{\xi}(\bar P',S'_z)$ , $\bar
u(\bar P,S_z) \bar P_{\mu}\bar P^\xi \gamma_{5}{u}_{\xi}(\bar
P',S'_z)$, and $\bar u(\bar P,S_z) g_{\mu}^
{\xi}\gamma_{5}{u}_{\xi}(\bar P',S'_z)$ to the right sides of
both Eq. (\ref{s2}) and Eq. (\ref{s3}) and solving for four algebraic equations, we can obtain analytical expressions of $g_{i}$($i=1,2,3,4$).
 % add some description of algebraic equations.

\subsubsection{the transition $\Xi_{cc}\to\Sigma^*_c (\Xi^*_c)$}
\begin{center}
\begin{figure}[htb]
\begin{tabular}{cc}
\scalebox{0.7}{\includegraphics{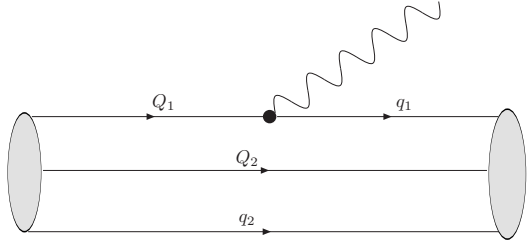}}
\end{tabular}
\caption{ The Feynman diagram for
 doubly charmed baryon weak decay, where $\bullet$ denotes
$V-A$ current vertex.
}\label{t2}
\end{figure}
\end{center}

For the weak decay of doubly charmed baryon $\Xi_{cc}\to\Sigma^*_c (\Xi^*_c)$,  one $c$ quark in the initial state $([cc] q_2)$ decays to a light quark $q_1$  in the final state  $(c [q_1 q_2])$ by weak interaction (Fig. \ref{t2}). This light quark $q_1$ combines with the light quark $q_2$  from the initial state to form a physical subsystem $q_1q_2$ which possesses definite spin
i.e. the quark composition  in the initial state is $[cc]q_2$, and that in the final state is $c[q_1 q_2]$ ($q_1$ and $ q_2$ may be same or not), which means neither the original [$cc$] nor the final [$q_1 q_2$] are spectators. The physical subsystem (diquark) is not spectator in the transition, so the quarks need to be rearranged. Therefore, the $[cc]$ diquark in the initial state needs to be destroyed and then rearranged with another light quark using a Racah transformation.
The related transformations are \cite{Zhao:2018zcb}
 \begin{eqnarray}
&&[c^1c^2]_1[q_2]=\frac{\sqrt{2}}{2}(-\frac{\sqrt{3}}{2}[c^2][c^1q_2]_0+\frac{1}{2}[c^2][c^1q_2]_1-\frac{\sqrt{3}}{2}[c^1][c^2q_2]_0+\frac{1}{2}[c^1][c^2q_2]_1),
 \end{eqnarray}
 where the subscript after brackets 0 or 1 denotes the total spin of two quarks in the brackets. The superscript for every $c$ quark is added to distinguish each other.
After the rearrangement $\Psi^{SS_z}_{S_{cc}}(\tilde{p}_i,\lambda_i)$ can be expressed to
 $-\frac{\sqrt{6}}{2}\Psi^{SS_z}_0(\tilde{p}_i,\lambda_i)+\frac{\sqrt{2}}{2}\Psi^{SS_z}_1(\tilde{p}_i,\lambda_i)$ where the subscript 0 and 1 are the total spin of $cq_2$ subsystem.

 Since the final state baryon $\mathcal B_f$ ($\Sigma^*_c$ or $\Xi^*_c$) has spin of $3/2$,   the Racah transformation leads to the following  rearrangement:
 \begin{eqnarray}
&&[c][q_1 q_2]_1=[q_1][c q_2]_{1}=[q_2][c q_1]_{1},
 \end{eqnarray}
and the expression $\Psi^{'SS_z}_{cq_{1}}(\tilde{p}'_i,\lambda'_i)$ or $\Psi^{'SS_z}_{cq_{2}}(\tilde{p}'_i,\lambda'_i)$ is just $\Psi^{'SS_z}_1(\tilde{p'}_i,\lambda'_i)$.

 % This is different from the Refs. \cite{Zhao:2018zcb,Hu:2022xzu}.

After the rearrangement the spectators in the process are isolated so the spectator approximation can be used in the calculation. In terms of the rearrangement the expressions of the form factors for the transition $\Xi_{cc}\to{\Sigma^{*}_c}(\Xi^{*}_c)$ have an additional factor $\frac{\sqrt{2}}{2}$ relative to those for $\Sigma_{b}\to{\Sigma^{*}_c}$.
% Why spectator approximation can be used in the calculation? Add some explanation

\section{Numerical Results}
\label{section_numerical}

\begin{table}
\caption{The  masses of the involved quarks and baryons (in units of
 GeV).}\label{Tab:t1}
\begin{ruledtabular}
\begin{tabular}{ccccccccccccc}
  $m_c$  & $m_b$  &$m_{d(u)}$  & $m_s$ & $\Sigma_{b}$ & $\Sigma_{c}^*$ &$\Omega_{b}$ &$\Omega_{c}^*$ &$\Xi_{b}^{'}$   &$\Xi_{cc}$
 &$\Xi_{c}^{*}$ \\\hline
  $1.3$   & $4.4$     & 0.25             &0.5              &5.811             &2.517            &6.065                &2.768                    &5.937               &3.621
     &2.646
\end{tabular}
\end{ruledtabular}
\end{table}

\begin{table}
\caption{The values of $\beta$ parameters (in units of  GeV)} \label{Tab:t2}
\begin{ruledtabular}
\begin{tabular}{ccccccccc}
  $\beta_{b[q_1q_2]}$  & $\beta_{c[q_1q_2]}$  &$\beta_{d[cq_2]}$  & $\beta_{s[cq_2]}$ & $\beta_{[c\bar c]}$ & $\beta_{[u\bar d]}$ &$\beta_{[s\bar s]}$ &$\beta_{[s\bar d]}$  & $\beta_{[c\bar q_2]}$ \\\hline
   $0.851$                 & $0.760$        & 0.656             &0.760                     & 0.655                              &0.263              &0.366                   &0.273     &0.381
\end{tabular}
\end{ruledtabular}
\end{table}

\subsection{The $\mathcal B_i(\frac{1}{2}^+)\to\mathcal B_f (\frac{3}{2}^+)$  form factors  }
In order to compute the relative transition rates of semi-leptonic decays and non-leptonic decays of $\mathcal B_i(\frac{1}{2}^+)\to\mathcal B_f (\frac{3}{2}^+)$ , we need to calculate the aforementioned form factors numerically. First of all, we need to fix the free parameters of the model, including masses of quarks and baryons, and wavefunction $\beta$ parameters.  The masses of quarks given in Ref. \cite{Cheng:2003sm} and the masses of baryons taken from Refs. \cite{LHCb:2017iph,ParticleDataGroup:2022pth} are listed in table \ref{Tab:t1}.

There is no precise measure of  the parameters $\beta_1$, $\beta_{23}$, $\beta_1'$ and $\beta_{23}'$ in the wave functions of the initial and the final baryons. Generally the reciprocal of $\beta$ is related to the electrical
radium of the baryon. Since the strong coupling strength between $q_1$ and
$q_2$ is half of that between $q_1\bar q_2$. Assuming  a Coulomb-like potential, one can expect the  radius of $q_1q_2$ to be $1/\sqrt{2}$ times that of $q_1\bar q_2$ i.e. $\beta_{q_1q_2}\approx\sqrt{2}\beta_{q_1\bar q_2}$. In
Ref. \cite{LeYaouanc:1988fx} in terms of the binding energy the
authors also obtained the same results.  Therefore in our work we use the $\beta$ values in the mesons case ($\beta_{q_1\bar q_2}$)
\cite{Cheng:2003sm} to estimate $\beta_{q_1 q_2}$. As for the value of $\beta_1$ we refer to the $\beta$ values of the heavy mesnons in Ref. \cite{Chang:2018zjq}.

Since the  form factors
$f_{i}$($i=1,2,3,4$) and $g_{i}$($i=1,2,3,4$) is calculated in the frame $q^+=0$
i.e. $q^2=-q^2_{\perp}\leq 0$ (the space-like region) one needs to
extend them into the time-like region to evaluate the transition
rates. In Refs. \cite {pentaquark1,pentaquark2} the form factors were parameterized using the
three-parameter form
\begin{eqnarray}\label{s145}
 F(q^2)=\frac{F(0)}{\left(1-\frac{q^2}{M_{\mathcal B_i}^2}\right)
  \left[1-a\left(\frac{q^2}{M_{\mathcal B_i}^2}\right)
  +b\left(\frac{q^2}{M_{\mathcal B_i}^2}\right)^2\right]}.
 \end{eqnarray}
However for the decay of doubly charmed baryon this parametric form doesn't work well, instead  a polynomial  form
was employed \cite{Ke:2019lcf}. So here we also use the three-parameter polynomial form
\begin{eqnarray}\label{s145p}
 F(q^2)=F(0)+a\frac{q^2}{M_{\mathcal B_i}^2}+b\left(\frac{q^2}{M_{\mathcal B_i}^2}\right)^2,
 \end{eqnarray}
where $F(q^2)$ represents the form factors $f_i$ and $g_i$.

Using the form factors calculated numerically in the space-like region we
 fit the parameters $a, b $ and $F(0)$ in the
un-physical region and then extrapolate  to the physical region with $q^2\geq 0$ through
 Eq. (\ref{s145p}).
\begin{table}
\caption{The $\Sigma_{b}\to \Sigma_c^*$ form factors given in the
  three-parameter form with $\beta_{23}= 2.9 \beta_{u\bar d}$.}\label{Tab:t23}
\begin{ruledtabular}
\begin{tabular}{cccc}
  $F$    &  $F(0)$ &  $a$  &  $b$ \\\hline
  $f_1$  &  -0.0140    &  0.0156   & 0.101  \\
$f_2$    &   0.130    &   0.364   & 0.408  \\
  $f_3$  &   -0.137   &   -0.310   &  -0.302  \\
  $f_4$  &    0.420   &   0.778   &  0.679  \\
  $g_1$  &  -0.309    & -0.740    &  -0.758 \\
  $g_2$  &   0.00908   &  0.0291   &  0.0400 \\
 $g_3$  &     0.197   & 0.481    & 0.502\\
  $g_4$  &   -0.744   & -1.60   & -1.55
\end{tabular}
\end{ruledtabular}
\end{table}

\begin{figure}[hhh]
\begin{center}
\scalebox{0.8}{\includegraphics{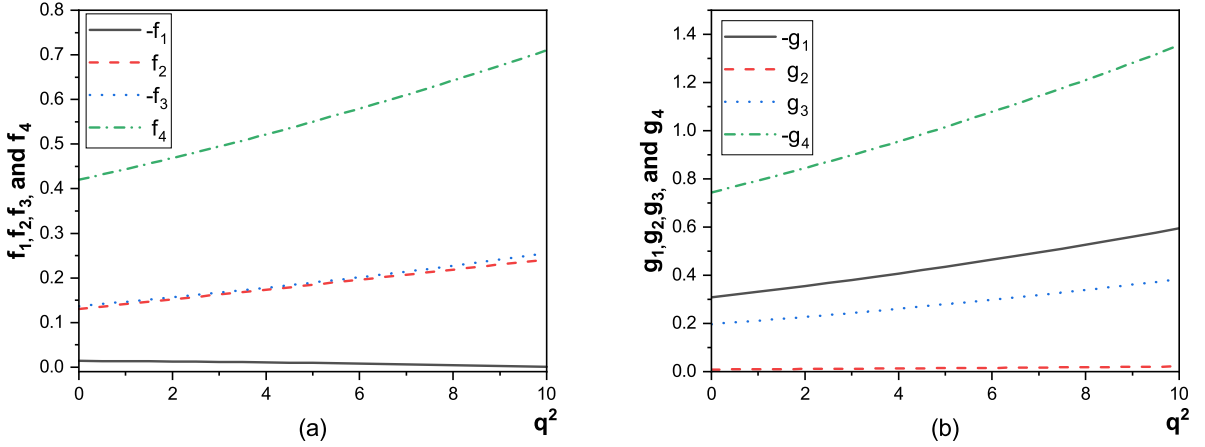}}
\end{center}
\caption{(a)  the form factors  $f_i\; (i=1,2,3,4)$and (b) the
form factors $g_i\; (i=1,2,3,4)$ with $\beta_{23}= 2.9
\beta_{u\bar d}$ } \label{f53}
\end{figure}

\subsubsection{the transition $\Sigma_b\to\Sigma^*_c$}
\label{form_factor_sigma}

Based on the forementioned discussion, we set
$\beta_1=\sqrt{2}\beta_{b\bar s}$ and $\beta_1'=
\sqrt{2}\beta_{c\bar s}$. However $u$ and $d$ quarks can be regarded as a $1^+$ diquark which means the distance
between $u$ and $d$ quarks should be smaller than normal case so $\beta_{23}$ will bigger than $\sqrt{2}\beta_{u\bar d}$. In Ref. \cite{Ke:2019smy} we fixed $\beta_{23}=\beta_{23}'=2.9 \beta_{u\bar d}$ and $\beta_{u\bar d}=0.263$ GeV \cite{Cheng:2003sm},  since $u$ and $d$ quarks are in the $^3S_1$ state.
The relevant values of $\beta$ are presented in table
\ref{Tab:t2}. With these parameters we calculate the
form factors and make theoretical predictions on the transition rates. The fitted values of $a,~b$ and $F(0)$ in the form
factors $f_{i}$ and $g_{i}$ are presented in
Table \ref{Tab:t23}.
The dependence of the form factors on $q^2$ is depicted in Fig. \ref{f53}.

From Fig. \ref{f53}, one can find that the
absolute values of the form factors $f_1(q^2)$ and $g_2(q^2)$ are
close to 0. The absolute values of $f_2(q^2)$
and $f_3(q^2)$ are almost close to each other at the small value of $q^2$ ($q^2<6$ Gev). Compared with the results in Ref. \cite{Ke:2017eqo},
the values of the $f_i$ and $g_i$ here
are similar to those obtained in Scheme II of Ref. \cite{Ke:2017eqo}
where the polarization of the $[ud]$ diquark depends on the momentum
of the diquark itself. Considering $1^+$ diquark is a so called bad diquark \cite{Wilczek:2004im}, which means the distance of $ud$ here is bigger than that for a $0^+$ diquark case, we estimate $\beta_{23}=\beta_{23}'< 2.9 \beta_{u\bar d}$
so we also set $\beta_{23}=\beta_{23}'= 2.0 \beta_{u\bar d}$ and $2.5 \beta_{u\bar d}$ to do the same calculation and compare the results on the semileptonic and nonleptonic decays later.

\subsubsection{the transition $\Xi_{cc}\to\Sigma^*_c(\Xi^*_c)$}

For the doubly baryon $\Xi_{cc}$, two $c$ quarks consist of a $1^+$ diquark so we set
$\beta_{1}=\beta_{c[cq_2]}= 2.9 \beta_{c\bar c}$ and $\beta_{23}=\beta_{23}'= \sqrt{2}\beta_{c\bar q_2}$. Since the rearrangement for $\Sigma^*_c(\Xi^*_c)$ we choose $\beta'_{1}=\beta_{d[cq_2]}=
\sqrt{2}\beta_{c\bar d}\, (\beta'_{1}=\beta_{s[cq_2]}=
\sqrt{2}\beta_{c\bar s})$. The relevant
parameters are collected in table \ref{Tab:t2}.

 The fitted values of $a,~b$ and $F(0)$ in the form factors $f_{i}$,
$g_{i}$ are presented in Table
\ref{Tab:t34}. The dependence of the
form factors on $q^2$ is depicted in
Fig. \ref{f63}. Compared with the curves of the form factors in Fig. \ref{f53}, those in Fig. \ref{f63} change
more significantly with $q^2$, especially when $q^2>2$ GeV.
Similarly we also set $\beta_{1}=\beta_{c[cq_2]}=2.0 \beta_{c\bar c}$ and $\beta_{1}=\beta_{c[cq_2]}= 2.5 \beta_{c\bar c}$ in the  calculation.

\begin{table}
\caption{The $\Xi_{cc}\to\Sigma_c^*$ form factors given
in the
  three-parameter form with  $\beta_{1}= 2.9 \beta_{c\bar c}$.}\label{Tab:t34}
\begin{ruledtabular}
\begin{tabular}{cccc}
   $F$    &  $F(0)$ &  $a$  &  $b$ \\\hline
  $f_1$  &  -0.0179    &  -0.0348   & -0.0226 \\
$f_2$    &   0.374    &   0.971  & 0.721  \\
  $f_3$  &   -0.381   &   -0.858  &  -0.603  \\
  $f_4$  &    0.429   &   0.552  &  0.302 \\
  $g_1$  &  -0.453   &  -0.933    &  -0.629 \\
  $g_2$  &        0     &  0   &  0\\
 $g_3$  &     0.320  &  0.683    &0.468\\
  $g_4$  &   -1.14  &  -1.90   & -1.18
\end{tabular}
\end{ruledtabular}
\end{table}
\begin{figure}[hhh]
\begin{center}
\scalebox{0.8}{\includegraphics{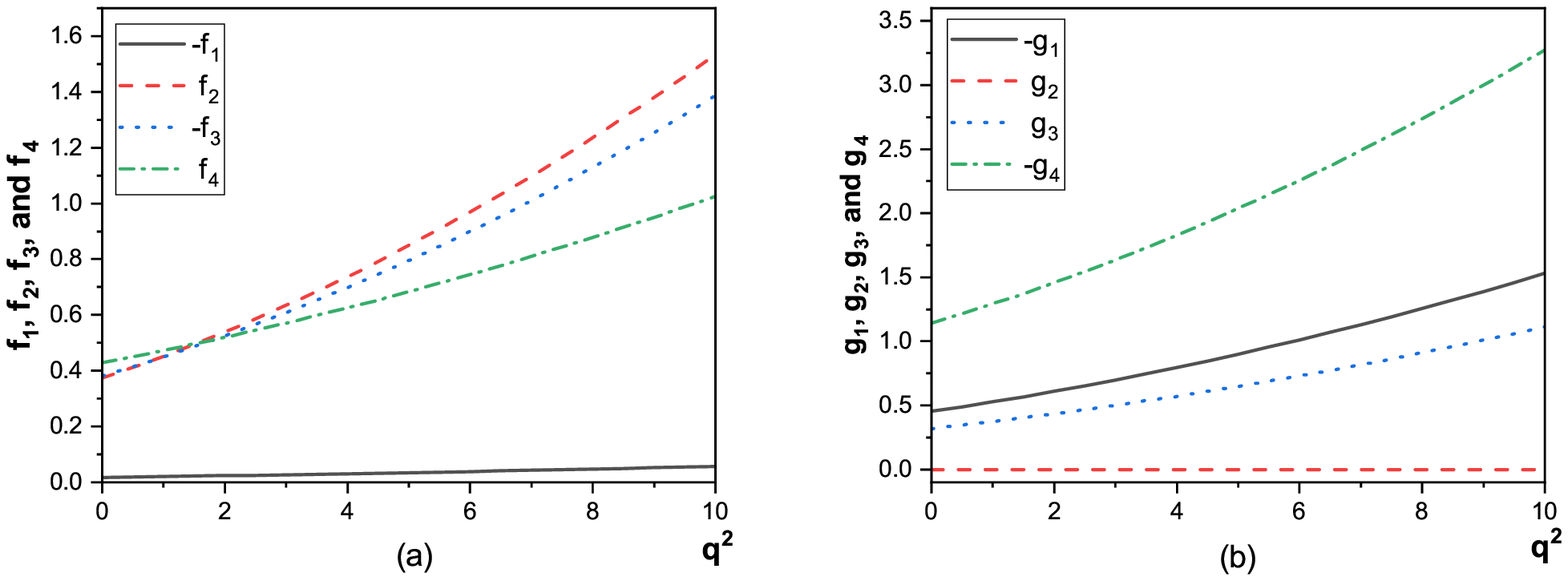}}
\end{center}
\caption{(a)  the form factors  $f_i\; (i=1,2,3,4)$ and (b) the
form factors $g_i\; (i=1,2,3,4)$ with $\beta_{1}= 2.9
\beta_{c\bar c}$ } \label{f63}
\end{figure}

\subsection{Semi-leptonic decay of $\mathcal B_i(\frac{1}{2}^+)\to\mathcal B_f (\frac{3}{2}^+) +l\bar{\nu}_l$}
\subsubsection{Semi-leptonic decay of single heavy baryon:  $\Sigma_b\to	\Sigma_c^* l\bar{\nu}_l$ }

Employing these form factors obtained in Sec. \ref{form_factor_sigma}, we evaluate the rates of $\Sigma_{b} \to \Sigma_{c}^*
l\bar{\nu}_l$.  At the same time we also depict the
differential decay rates of the $\Sigma_b \to \Sigma_c^*
+l\bar{\nu}_l$ which depend on $\omega$ ( definition can be found in Appendix B) in Fig. \ref{f54}. We calculate the total decay widths, longitudinal
decay widths, transverse decay widths and the ratio of the
longitudinal to transverse decay rates $R$ with
$\beta_{23}=\beta_{23}'=2.0 \beta_{u\bar d}$,
$2.5 \beta_{u\bar d}$, and
$2.9 \beta_{u\bar d}$, respectively.
 The results are listed in table \ref{Tab:t4}.

As we can see in table \ref{Tab:t4},  the width increases slowly when
the value of $\beta_{23}$ decreases and the
decay width is a little smaller than that in Ref. \cite{Ke:2017eqo} where quark-diquark picture for baryons within the LFQM was employed. The
longitudinal decay rate, transverse decay rate and their ratio $R$ are also a little difference under the two types of  models of baryon.
In Ref. \cite{Ivanov:1996fj} the authors employed
relativistic three-quark model to calculate theses form factor and their
predictions on the decay width of $\Sigma_b \to\Sigma_c^* +l\bar{\nu}_l$ are almost
twice larger than this work. In summary,  our results on the decay width of $\Sigma_b \to\Sigma_c^* +l\bar{\nu}_l$  are less than the
results in Refs. \cite{Ebert:2006rp,Ivanov:1996fj,Ivanov:1998ya,Ke:2017eqo}, which means a small $\beta_{23}$ ($\beta_{23}'$) is more reasonable if we want to make the prediction close to those in references.

We also use the same method to calculate the  decay width of
$\Omega_{b}\to\Omega_{c}^* l\bar{\nu}_l$, $\Xi_{b}^{'}\to\Xi_{c}^*
l\bar{\nu}_l$ with $\beta_{23}=\beta_{23}'= 2.0 \beta_{q_1\bar
q_2}$ and the results are shown in table \ref{Tab:t5}. From
table \ref{Tab:t5}, we can see that the widths of $\Sigma_{b} \to
\Sigma_{c}^* l\bar{\nu}_l, \Omega_{b}\to\Omega_{c}^*
l\bar{\nu}_l$ and $\Xi_{b}^{'}\to\Xi_{c}^* l\bar{\nu}_l$ are close to each other.
Similarly, they are lower than those in the
Ref. \cite{Ebert:2006rp}.

\subsubsection{Semi-leptonic decay of doubly charmed baryon: $\Xi_{cc}\to\Sigma_{c}^{*} l \bar{\nu}_l$ and	$\Xi_{cc}\to\Xi_{c}^{*} l \bar{\nu}_l$}

For the weak decay of doubly charmed baryon, we calculate the
decay rates of $\Xi_{cc}\to\Sigma_{c}^{*} l \bar{\nu}_l$ and
$\Xi_{cc}\to\Xi_{c}^{*} l \bar{\nu}_l$. The
curves of the differential decay widths depending on $\omega$ for $\Xi_{cc}\to\Sigma_{c}^{*} l \bar{\nu}_l$ are depicted in Fig. \ref{f55} which is very similar to that for $\Sigma_b \to \Sigma_c^*
+l\bar{\nu}_l$.  The
curves for $\Xi_{cc}\to\Xi^*_{c} l \bar{\nu}_l$ are similar to those in Fig. \ref{f55}  except their peak values are 20 times bigger than those for $\Xi_{cc}\to\Sigma_{c}^{*} l \bar{\nu}_l$ so we omit the figure. The numerical results with
$\beta_{1}= 2.0 \beta_{c\bar c}$, $2.5
\beta_{c\bar c}$, $2.9 \beta_{c\bar c}$ are presented in \ref{Tab:t52} and
 \ref{Tab:t54}, respectively. One can find that the differential decay rate
of $\Xi_{cc}\to\Sigma_{c}^{*} (\Xi_{c}^{*}) l \bar{\nu}_l$ increases with the decrease of the $\beta_{1}$ value
and the differential decay rate with $\beta_{1}= 2.0 \beta_{c\bar c}$ is close to that in
Refs. \cite{Zhao:2018zcb, Hu:2020mxk} where the
decay was explored with the quark-diquark picture in the
LFQM.
When $\beta_{1}= 2.9\beta_{c\bar c}$, the result of
$\Xi_{cc}\to\Sigma_{c}^{*} l \bar{\nu}_l$ is about half of the
Refs. \cite{Zhao:2018zcb, Hu:2020mxk}. Our results here indicate $\beta_{1}$ prefers a small number, such as  $2.0\beta_{c\bar c}$ if the predictions in references are accurate.

\begin{figure}[hhh]
\begin{center}
\scalebox{0.8}{\includegraphics{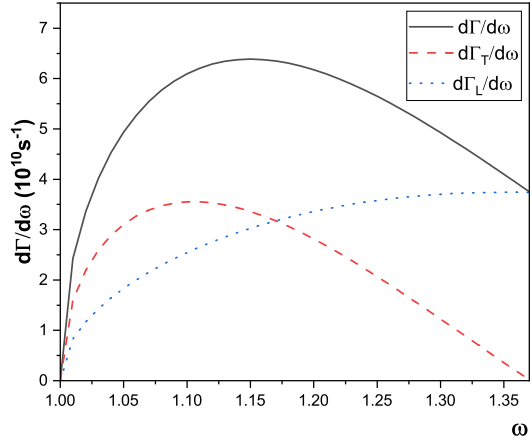}}
\end{center}
\caption{ Differential decay rates $d\Gamma/d\omega$ for the decay
$\Sigma_b \to \Sigma_c^* l\bar{\nu}_l$ with $\beta_{23}= 2.9 \beta_{u\bar d}$} \label{f54}
\end{figure}

\begin{figure}[hhh]
\begin{center}
\scalebox{0.8}{\includegraphics{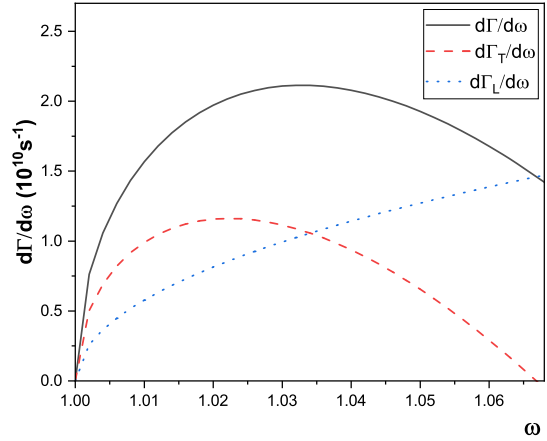}}
\end{center}
\caption{ Differential decay rates $d\Gamma/d\omega$ for the decay
$\Xi_{cc}\to\Sigma_{c}^{*} l \bar{\nu}_l$ with $\beta_{1}= 2.9 \beta_{c\bar c}$} \label{f55}
\end{figure}

\begin{table}
\caption{The widths and polarization asymmetries of $\Sigma_b\to
\Sigma_c^* l\bar{\nu}_l$ .}\label{Tab:t4}
\begin{ruledtabular}
\begin{tabular}{c|ccccc}
 &  $\Gamma$ ($10^{10}{\rm s}^{-1}$) &  $\Gamma_L$  &  $\Gamma_T$   & $R$     \\\hline
  this work($\beta_{23}= 2.0 \beta_{u\bar d}$)& 2.12&1.17 &0.953 &1.23  \\\hline
  this work($\beta_{23}= 2.5 \beta_{u\bar d}$)& 2.01 &1.11 &0.898 &1.24  \\\hline
  this work($\beta_{23}= 2.9 \beta_{u\bar d}$)& 1.95 &1.09 &0.860 &1.27 \\\hline
 the results in Ref. \cite{Ebert:2006rp} &   3.23      & 1.61  & 1.62  & 0.99\\\hline
 the results in Ref. \cite{Ivanov:1996fj}& 4.56   &  2.49 & 2.07 & 1.20 \\\hline
  the results in Ref. \cite{Ivanov:1998ya}&3.75 & - &-&- \\\hline
  the results in Ref. \cite{Ke:2017eqo}
    &  3.17$\pm$0.30   & 1.58$\pm$0.16  & 1.59$\pm$0.13& 0.994$\pm0.024$
\end{tabular}
\end{ruledtabular}
\end{table}

\begin{table}
\caption{ The widths and polarization asymmetries of the semileptonic decay between two single heavy
baryons ($\beta_{23}= 2.0 \beta_{u\bar d}$). }\label{Tab:t5}
\begin{ruledtabular}
\begin{tabular}{c|cccc|cccc}
 &  $\Gamma$ ($10^{10}{\rm s}^{-1}$)          &  $\Gamma_L$  &  $\Gamma_T$   & $R$  &  $\Gamma$ ($10^{10}{\rm s}^{-1}$) \cite{Ebert:2006rp}         &  $\Gamma_L$  &  $\Gamma_T$   & $R$    \\\hline
 $\Sigma_b\to\Sigma_c^* l\bar{\nu}_l$          & 2.12&1.17 &0.953&1.23   & 3.23&1.61&1.62&0.99 \\\hline
 $\Omega_{b}\to\Omega_{c}^*l\bar{\nu}_l$   & 1.99&1.08&0.913&1.18 & 3.09&1.52&1.57&0.97 \\\hline
 $\Xi_{b}^{'}\to\Xi_{c}^*l\bar{\nu}_l$               & 2.08&1.13&0.947&1.19    & 3.03&1.48&1.55&0.95
\end{tabular}
\end{ruledtabular}
\end{table}

\begin{table}
\caption{ The widths and polarization asymmetries of the transition
$\Xi_{cc}\to\Sigma_{c}^{*} l \bar{\nu}_l$. }\label{Tab:t52}
\begin{ruledtabular}
\begin{tabular}{c|ccccc}
 &  $\Gamma$ ($10^{10}{\rm s}^{-1}$)            &   $\Gamma_L$            & $\Gamma_T$             & $R$     \\\hline
  this work($\beta_{1}= 2.0 \beta_{c\bar c}$)& 0.181    &0.100            &0.0807                         &1.24  \\\hline
  this work($\beta_{1}= 2.5 \beta_{c\bar c}$)& 0.145  &0.0802           &0.0652                          &1.23  \\\hline
  this work($\beta_{1}= 2.9 \beta_{c\bar c}$)& 0.120  &0.0658            & 0.0537                        &1.23 \\\hline
  the results in \cite{Zhao:2018zcb} & 0.191    & -   & - & -\\\hline
  the results in  \cite{Hu:2020mxk}& 0.217    &  - & - & -
\end{tabular}
\end{ruledtabular}
\end{table}

\begin{table}
\caption{ The widths and polarization asymmetries
of the transition $\Xi_{cc}\to\Xi_{c}^{*} l \bar{\nu}_l$.}\label{Tab:t54}
\begin{ruledtabular}
\begin{tabular}{c|ccccc}
 &  $\Gamma$ ($10^{10}{\rm s}^{-1}$)                                          &   $\Gamma_L$            & $\Gamma_T$                     & $R$     \\\hline
  this work($\beta_{1}= 2.0 \beta_{c\bar c}$)&  2.44    &1.18             &1.26                         &0.937 \\\hline
  this work($\beta_{1}= 2.5 \beta_{c\bar c}$)&  2.02    &0.991             &1.03                         &0.962  \\\hline
  this work($\beta_{1}=2.9 \beta_{c\bar c}$)& 1.67   &0.831            &0.841                         &0.988 \\\hline
  the results in  \cite{Zhao:2018zcb} &   2.39    & -   & - & -\\\hline
the results in   \cite{Hu:2020mxk}& 2.64    &  - & - & -
\end{tabular}
\end{ruledtabular}
\end{table}

\subsection{Non-leptonic decays of $\mathcal B_i(\frac{1}{2}^+)\to\mathcal B_f (\frac{3}{2}^+)+ M$}

Because of the strong interaction, the non-leptonic
decays are more complicated than the semi-leptonic processes.
Here we adopt the theoretical framework of factorization assumption,
where the hadronic transition matrix element can be factorized into a
product of two independent matrix elements of currents.

% Add some detail or reference of fractorization assumption. Maybe also explain why this assumption is valid.

\subsubsection{Non-leptonic decays of single heavy baryon:
	$\Sigma_b\to	\Sigma_c^* + M $ }
 For $b\to c$ transition, the hadronic transition
matrix element is
\begin{eqnarray}\label{s147}
&& \la \mathcal B_f(\frac{3}{2}^+)M \mid \mathcal{H} \mid \mathcal B_i(\frac{1}{2}^+) \ra  \nonumber \\
 &=&\frac{G_FV_{bc}V^*_{q_aq_b}}{\sqrt{2}}\la M \mid
\bar{q_b} \gamma^{\mu} (1-\gamma_{5}) q_a \mid 0\ra\la \mathcal B_f
(\frac{3}{2}^+) \mid \bar{c} \gamma^{\mu} (1-\gamma_{5}) b \mid
\mathcal B_i(\frac{1}{2}^+) \ra,
\end{eqnarray}
where the term $\la M \mid \bar{q}_b \gamma^{\mu}
(1-\gamma_{5}) q_a \mid 0\ra$ can be written as the decay constant
of meson $M$ (where $q_a$ and $q_b$ denote heavy or light quark flavors) and the second one $\la \mathcal B_f
(\frac{3}{2}^+)\mid \bar{c} \gamma^{\mu} (1-\gamma_{5}) b
\mid\mathcal B_i(\frac{1}{2}^+) \ra$ is determined by the
form factors we obtained. The Fermi constant and CKM matrix elements are selected
from Ref.\cite{ParticleDataGroup:2022pth}
\beq\label{s148}
& G_F=1.1664\times 10^{-5} {\rm GeV}^{-2}, \nonumber \\
& V_{cb}=0.0416,  \qquad V_{ud}=0.9738,  \qquad V_{us}=0.2257,   \qquad V_{cd}=0.230,   \qquad V_{cs}=0.957\nonumber.
 \eeq

In table \ref{Tab:t6}, we present the results of the main two-body
decay channels for $\Sigma_{b}$ with the different value of
$\beta_{23}(\beta_{23}')$. The decay constants are
derived from the Ref. \cite{Cheng:2003sm}. From the table \ref{Tab:t6}
one also can notice that the widths increase  with the decrease of the
value of $\beta_{23}\,(\beta_{23}')$ and
 the results with $\beta_{23}=\beta_{23}'= 2.0 \beta_{u\bar d}$ are close to those with the heavy
quark limit in Ref.\cite{Ke:2017eqo}.

\subsubsection{Non-leptonic decays of doubly charmed baryon: $\Xi_{cc}\to\Sigma^*_c(\Xi^*_c)+ M$}
The non-leptonic decays results of the doubly charmed baryon are
listed in table \ref{Tab:t8}. As we can see from the table that
the non-leptonic decay widths increase significently with the decrease of
the value of $\beta_{1}$. A smaller value of $\beta_{1}$ is more consistent with results in previous studies, such as Ref.\cite{Zhao:2018zcb}.
Future experiments are needed to constrain the $\beta$ parameters and test our model predictions.
% Experimental results? current status of experiements

%More data are needed to determine the parameter, and then we can do more precise predictions.

%Compared with the results in Ref.\cite{Zhao:2018zcb},
% for the results also prefers a small value.
%More data are needed to determine the parameter, and then we can do more precise predictions.

\begin{table}
\caption{The widths (in unit $10^{10}{\rm s}^{-1}$) of the non-leptonic
decays $\Sigma_b\to\Sigma_c^* + M$ .}\label{Tab:t6}
\begin{ruledtabular}
\begin{tabular}{ccccc}
mode&  $\Gamma (\beta_{23}=2.0 \beta_{u\bar d})$  &  $\Gamma (\beta_{23}=2.5
\beta_{u\bar d})$ & $\Gamma (\beta_{23}=2.9 \beta_{u\bar d})$ & $\Gamma $\cite{Ke:2017eqo}  \\\hline
  $\Sigma_b^0\to\Sigma_c^* \pi^-$ & 0.136 & 0.127 & 0.123 & $0.132\pm0.025$ \\\hline
  $\Sigma_b^0\to\Sigma_c^* \rho^-$ & 0.405 & 0.378 & 0.366 & $0.411\pm0.082$\\\hline
  $\Sigma_b^0\to\Sigma_c^* K^-$  & 0.0107  &0.00998 &0.00968 & $0.0104\pm0.0019$  \\\hline
  $\Sigma_b^0\to\Sigma_c^* K^{*-}$ & 0.0209 & 0.0195 & 0.0189  & $0.0217\pm0.0039$\\\hline
  $\Sigma_b^0\to\Sigma_c^* a_1^-$ & 0.403 & 0.376 & 0.365  & $0.437\pm0.076$\\\hline
  $\Sigma_b^0\to\Sigma_c^* D^-$  & 0.0127 &0.0121& 0.0122  & $0.0139\pm0.0020$ \\\hline
  $\Sigma_b^0\to\Sigma_c^* D^{*-}$ & 0.0306 &0.0288 & 0.0280 & $0.0431\pm0.0058$ \\\hline
  $\Sigma_b^0\to\Sigma_c^* {D_s^-}$ & 0.313&0.299 & 0.302  & $0.351\pm0.0048$ \\\hline
  $\Sigma_b^0\to\Sigma_c^* D^{*-}_s$ & 0.667 &0.628 & 0.612 & $0.990\pm0.13$
\end{tabular}
\end{ruledtabular}
\end{table}

\begin{table}
\caption{ The nonleptonic decay widths (in unit $10^{10}{\rm s}^{-1}$) of
 the doubly charmed baryon to the single charmed baryon.}\label{Tab:t8}
\begin{ruledtabular}
\begin{tabular}{cccccc}
 &$\Gamma (\beta_{1}=2.0 \beta_{c\bar c})$ &$\Gamma (\beta_{1}=2.5 \beta_{c\bar c})$
  &$\Gamma (\beta_{1}=2.9 \beta_{c\bar c})$ & $\Gamma $ \cite{Zhao:2018zcb}\\\hline
  $\Xi_{cc}\to\Sigma_{c}^{*}  \pi$ & 0.116  & 0.0924  &0.0758 & 0.176  \\\hline
  $\Xi_{cc}\to\Sigma_{c}^{*}  \rho$ &0.459& 0.369 &0.304 & 0.574  \\\hline
  $\Xi_{cc}\to\Sigma_{c}^{*}  K$ & $7.34\times 10^{-3}$ & $5.77\times 10^{-3}$&$4.70\times 10^{-3}$  & $7.55\times 10^{-3}$   \\\hline
  $\Xi_{cc}\to\Sigma_{c}^{*} K^{*}$ &$2.22\times 10^{-2}$ & $1.79\times 10^{-2}$&$1.47\times 10^{-2}$& $2.43\times 10^{-2}$   \\\hline
 $\Xi_{cc}\to\Xi_{c}^{*} \pi$        & $2.39$  & $1.91$  &$1.55$& $3.39$  \\\hline
  $\Xi_{cc}\to\Xi_{c}^{*}  \rho$   &$9.03$ & $7.33$ &$5.98$ & $7.07$  \\\hline
  $\Xi_{cc}\to\Xi_{c}^{*}K $        &0.135 & 0.106 &0.0857 & 0.106   \\\hline
  $\Xi_{cc}\to\Xi_{c}^{*}K^{*}$   &0.355 & 0.288 &0.235 & 0.190
\end{tabular}
\end{ruledtabular}
\end{table}

\section{Conclusions and discussions}

In this paper, we study the  weak decays between two heavy baryons $\mathcal{B}_i(\frac{1}{2}^+)\to  \mathcal{B}_f(\frac{3}{2}^+)$ with the three-quark picture of baryon in the LFQM. We derive the general form of transition amplitude, and obtain analytical expression of form factors for specific transition processes: $\Sigma_{b}\to\Sigma_{c}^*$ and $\Xi_{cc}\to\Sigma^*_c(\Xi^*_c)$ . For weak decay of $\Sigma_{b}\to\Sigma_{c}^*$, the $b$ quark decays to $c$ quark and $ud$ subsystem with definite spin can be regarded as spectator in initial and final states. For  the transition $\Xi_{cc}\to\Sigma^*_c(\Xi^*_c)$, the $cc$ system in initial state and the $ud$ ($su$ or $sd$)  system in final state possess definite spins, but they are not spectators. In that case,  quark rearrangement is adopted in our calculation of form factors. We then compute numerical values of these form factors with reasonable assumptions of model parameters. Last, we calculate rate of semi-leptonic and non-leptonic decays  of $\Sigma_{b}$ ($\Sigma_b\to	\Sigma_c^* l\bar{\nu}_l$,  $\Sigma_b\to	\Sigma_c^*  + M$) and $\Xi_{cc}$ ( $\Xi_{cc}\to\Sigma_{c}^{*}(\Xi_{c}^{*}) l \bar{\nu}_l$, $\Xi_{cc}\to\Sigma_{c}^{*} (\Xi_{c}^{*}) + M$) based on numerical results of these form factors.

The weak decays
of $\Sigma_{b}\to\Sigma_{c}^*$ were studied in Ref. \cite{Ke:2017eqo} where the quark-diquark picture  was employed  in
the LFQM. Instead of  quark-diquark picture, here we use the three-quark picture to revisit the transition of $\Sigma_{b}\to\Sigma_{c}^*$. During the transition, the two light quarks serve as the spectators and maintain their all quantum numbers (spin, color). The associated momentum is also unchanged. The  $b$ quark in initial state transits to $c$ quark by emitting two leptons mediated by gauge bosons $W^{\pm}$.

We calculate the form factors  for the transition
$\Sigma_{b}\to\Sigma_{c}^*$. The values of the $f_i$ and $g_i$
are similar to those obtained in Scheme II of Ref. \cite{Ke:2017eqo}
where the polarization of the $[ud]$ diquark depends on the momentum
of the diquark itself. Considering $1^+$ diquark is a so-called bad diquark, i.e. the distance between the two quarks is larger than a good diquark case, we estimate $\beta_{23}=\beta_{23}'< 2.9 \beta_{u\bar d}$
so we also set $\beta_{23}=\beta_{23}'= 2.0 \beta_{u\bar d}$ and $2.5 \beta_{u\bar d}$ to perform the same calculation.
% Please rewrite this paragraph. It is not well wroten. Explain "Scheme II" and "bad diquark". Motivation to consider a bad diquark.
%
With computed form factors we evaluate the decay widths of the
semi-leptonic $\Sigma_b\to\Sigma_c^* l\bar{\nu}_l$ and
non-leptonic $\Sigma_b\to\Sigma_c^* +M$ with different values of
 $\beta_{23}$. Comparing the results in other references, we find $\beta_{23}$ and $\beta_{23}'$ prefer a small value if we want our predictions close to those in the references. We also compute the  semi-leptonic decay
width of $\Omega_{b}\to\Omega_{c}^* l\bar{\nu}_l$ and
$\Xi_{b}^{'}\to\Xi_{c}^* l\bar{\nu}_l$ with
$\beta_{23}=\beta_{23}'=2.0 \beta_{q_1\bar q_2}$.

For the weak decay of the doubly charmed
baryon $\Xi_{cc}$, a heavy quark $c$ in the initial state decays to a
light quark ($s$ or $d$) in the final state through the weak interaction. However the $cc$ pair in initial state and $ud$ ($us$ or $ds$) in the final state possess definite spin which can be regarded as diquarks.
In this way, neither the initial physical
diquark $cc$ nor the final physical diquark $ud$ ($us$ or $ds$) are
spectators. Therefore, the three-quark picture is more suitable here. We have rearranged the quarks in the initial
and final states using the Racah transformation so the effective spectator can be isolated from the baryon. We calculate
the form factors in the space-like region and then
extend them to the time-like region (the physical region) by using
the three-parameter form. Using these form factors we calculate the widths
of semi-leptonic decay $\Xi_{cc}\to\Sigma^*_c(\Xi^*_c)  l \bar{\nu}_l$ and
non-leptonic decay $\Xi_{cc}\to\Sigma^*_c(\Xi^*_c)+M$, respectively.
For the weak decay of $\Xi_{cc}\to\Sigma^*_c(\Xi^*_c)  l \bar{\nu}_l$,
the decay width increases when the $\beta_{1}$ value decreases.
Our results on the semi-leptonic decay $\Xi_{cc}\to\Sigma_{c}^{*} (\Xi_{c}^{*} l \bar{\nu}_l) l \bar{\nu}_l$
with $\beta_{1}=2.0 \beta_{c\bar c}$ are close to the
Ref. \cite{Zhao:2018zcb}. Future experiments will be necessary to precisely determine the parameters and test model predictions.

\section*{Acknowledgement}

 This work is supported by the National Natural Science Foundation
of China (NNSFC) under the Contracts No. 12075167, 11975165 and 12235018.

\appendix

\section{the form factor of
 $\mathcal B_i(\frac{1}{2}^+)\to\mathcal B_f (\frac{3}{2}^+)$ }
$\bar u(\bar P,S_z)\gamma_{\mu}\bar P^\xi {u}_{\xi}(\bar P',S'_z)$,
$\bar u(\bar P,S_z)\bar P'_{\mu}\bar P^\xi{u}_{\xi}(\bar
P',S'_z)$, $\bar u(\bar P,S_z) \bar P_{\mu}\bar
P^\xi{u}_{\xi}(\bar P',S'_z)$, $\bar u(\bar P,S_z) g_{\mu}^{
\xi}{u}_{\xi}(\bar P',S'_z)$ are multiplied to the right side of
Eq.(\ref{s3}), and then we have
\begin{eqnarray}\label{s01}
  F_1&=&\int\frac{ d x_2 d^2 k^2_{2\perp}}{2(2\pi)^3}\frac{ d x_3 d^2 k^2_{3\perp}}{2(2\pi)^3}\frac{{\phi_{\mathcal B_f}^*(x',k'_{\perp})
  \phi_{\mathcal B_i}(x,k_{\perp})}Tr[\gamma_{\perp}^\alpha(\bar{P'}\!\!\!\!\!\slash+M_0')(p_3\!\!\!\!\!\slash+m_3)
  (\bar{P}\!\!\!\!\!\slash+M_0)\gamma_{\perp}^\beta(p_2\!\!\!\!\!\slash-m_2)]}{16\sqrt{6x_1x'_1M_0^3M_0'^3(m_1+e_1)
 (m_2+e_2)(m_3+e_3)(m_1'+e_1')
 (m_2'+e_2')(m_3'+e_3')}} \nonumber \\
   &&\times\sum_{S_z,S'_z}{\rm Tr}[{u}_{\xi}(\bar P',S'_z)\bar{u}_{\alpha}(\bar{P'},S'_z)
  (p_1\!\!\!\!\!\slash'+m'_1)\gamma^{\mu}\gamma_{5}
  (p_1\!\!\!\!\!\slash+m_1)\gamma_{\perp\beta}\gamma_{5}  u(\bar{P},S_z)\bar u(\bar P,S_z)\gamma_{\mu}\bar P^\xi],\\
   F_2&=&\int\frac{ d x_2 d^2 k^2_{2\perp}}{2(2\pi)^3}\frac{ d x_3 d^2 k^2_{3\perp}}{2(2\pi)^3}\frac{{\phi_{\mathcal B_f}^*(x',k'_{\perp})
  \phi_{\mathcal B_i}(x,k_{\perp})}Tr[\gamma_{\perp}^\alpha(\bar{P'}\!\!\!\!\!\slash+M_0')(p_3\!\!\!\!\!\slash+m_3)
  (\bar{P}\!\!\!\!\!\slash+M_0)\gamma_{\perp}^\beta(p_2\!\!\!\!\!\slash-m_2)]}{16\sqrt{6x_1x'_1M_0^3M_0'^3(m_1+e_1)
 (m_2+e_2)(m_3+e_3)(m_1'+e_1')
 (m_2'+e_2')(m_3'+e_3')}} \nonumber \\
   &&\times\sum_{S_z,S'_z}{\rm Tr}[{u}_{\xi}(\bar P',S'_z)\bar{u}_{\alpha}(\bar{P'},S'_z)
  (p_1\!\!\!\!\!\slash'+m'_1)\gamma^{\mu}\gamma_{5}
  (p_1\!\!\!\!\!\slash+m_1)\gamma_{\perp\beta}\gamma_{5}  u(\bar{P},S_z)\bar u(\bar P,S_z)\bar P'_{\mu}\bar
P^\xi],\\
  F_3&=&\int\frac{ d x_2 d^2 k^2_{2\perp}}{2(2\pi)^3}\frac{ d x_3 d^2 k^2_{3\perp}}{2(2\pi)^3}\frac{{\phi_{\mathcal B_f}^*(x',k'_{\perp})
  \phi_{\mathcal B_i}(x,k_{\perp})}Tr[\gamma_{\perp}^\alpha(\bar{P'}\!\!\!\!\!\slash+M_0')(p_3\!\!\!\!\!\slash+m_3)
  (\bar{P}\!\!\!\!\!\slash+M_0)\gamma_{\perp}^\beta(p_2\!\!\!\!\!\slash-m_2)]}{16\sqrt{6x_1x'_1M_0^3M_0'^3(m_1+e_1)
 (m_2+e_2)(m_3+e_3)(m_1'+e_1')
 (m_2'+e_2')(m_3'+e_3')}} \nonumber \\
   &&\times\sum_{S_z,S'_z}{\rm Tr}[{u}_{\xi}(\bar P',S'_z)\bar{u}_{\alpha}(\bar{P'},S'_z)
  (p_1\!\!\!\!\!\slash'+m'_1)\gamma^{\mu}\gamma_{5}
  (p_1\!\!\!\!\!\slash+m_1)\gamma_{\perp\beta}\gamma_{5}  u(\bar{P},S_z)\bar u(\bar P,S_z)\bar
P_{\mu}\bar P^\xi],\\
   F_4&=&\int\frac{ d x_2 d^2 k^2_{2\perp}}{2(2\pi)^3}\frac{ d x_3 d^2 k^2_{3\perp}}{2(2\pi)^3}\frac{{\phi_{\mathcal B_f}^*(x',k'_{\perp})
  \phi_{\mathcal B_i}(x,k_{\perp})}Tr[\gamma_{\perp}^\alpha(\bar{P'}\!\!\!\!\!\slash+M_0')(p_3\!\!\!\!\!\slash+m_3)
  (\bar{P}\!\!\!\!\!\slash+M_0)\gamma_{\perp}^\beta(p_2\!\!\!\!\!\slash-m_2)]}{16\sqrt{6x_1x'_1M_0^3M_0'^3(m_1+e_1)
 (m_2+e_2)(m_3+e_3)(m_1'+e_1')
 (m_2'+e_2')(m_3'+e_3')}} \nonumber \\
   &&\times\sum_{S_z,S'_z}{\rm Tr}[{u}_{\xi}(\bar P',S'_z)\bar{u}_{\alpha}(\bar{P'},S'_z)
  (p_1\!\!\!\!\!\slash'+m'_1)\gamma^{\mu}\gamma_{5}
  (p_1\!\!\!\!\!\slash+m_1)\gamma_{\perp\beta}\gamma_{5}  u(\bar{P},S_z)\bar u(\bar P,S_z)g_{\mu}^{
\xi}] .
\end{eqnarray}

Simultaneously, $\bar u(\bar P,S_z)\gamma_{\mu}\bar P^\xi {u}_{\xi}(\bar P',S'_z)$,
$\bar u(\bar P,S_z)\bar P'_{\mu}\bar P^\xi{u}_{\xi}(\bar
P',S'_z)$, $\bar u(\bar P,S_z) \bar P_{\mu}\bar
P^\xi{u}_{\xi}(\bar P',S'_z)$, $\bar u(\bar P,S_z) g_{\mu}^{
\xi}{u}_{\xi}(\bar P',S'_z)$ are multiplied to the right side of Eq.(\ref{s2}), one can obtain
\begin{eqnarray}\label{s21}
F_1 &=&{\rm Tr}\{{u}_{\xi}(\bar P',S'_z) \bar{u}_{\alpha}(\bar
P',S'_z) \left[ \gamma^{\mu}\bar P^\alpha
\frac{f_{1}(q^{2})}{M_{\mathcal B_i}}
 + \frac{f_{2}(q^{2})}{M^2_{\mathcal B_i}}\bar P^\alpha \bar P^\mu+\frac{f_{3}(q^{2})}{M_{\mathcal B_i}M_{\mathcal B_f}}\bar P^\alpha \bar P'^\mu
 +f_4g^{\alpha \mu}
 \right] \nonumber\\&&u(\bar P,S_z)\bar u(\bar{P},S_z)\gamma_{\mu}\bar P^\xi \},\\F_2 &=&{\rm Tr}\{{u}_{\xi}(\bar
P',S'_z) \bar{u}_{\alpha}(\bar P',S'_z) \left[\gamma^{\mu}\bar
P^\alpha \frac{f_{1}(q^{2})}{M_{\mathcal B_i}}
 + \frac{f_{2}(q^{2})}{M^2_{\mathcal B_i}}\bar P^\alpha \bar P^\mu+\frac{f_{3}(q^{2})}{M_{\mathcal B_i}M_{\mathcal B_f}}\bar P^\alpha \bar P'^\mu
 +f_4g^{\alpha \mu}
 \right] \nonumber\\&&u(\bar P,S_z)\bar u(\bar{P},S_z)\bar P'_{\mu}\bar
P^\xi \},\\F_3 &=&{\rm Tr}\{{u}_{\xi}(\bar P',S'_z)
\bar{u}_{\alpha}(\bar P',S'_z) \left[\gamma^{\mu}\bar P^\alpha
\frac{f_{1}(q^{2})}{M_{\mathcal B_i}}
 + \frac{f_{2}(q^{2})}{M^2_{\mathcal B_i}}\bar P^\alpha \bar P^\mu+\frac{f_{3}(q^{2})}{M_{\mathcal B_i}M_{\mathcal B_f}}\bar P^\alpha \bar P'^\mu
 +f_4g^{\alpha \mu}
 \right] \nonumber\\&&u(\bar P,S_z)\bar u(\bar{P},S_z)\bar
P_{\mu}\bar P^\xi \},
  \\F_4 &=&{\rm Tr}\{{u}_{\xi}(\bar
P',S'_z) \bar{u}_{\alpha}(\bar P',S'_z) \left[ \gamma^{\mu}\bar
P^\alpha \frac{f_{1}(q^{2})}{M_{\mathcal B_i}}
 + \frac{f_{2}(q^{2})}{M^2_{\mathcal B_i}}\bar P^\alpha \bar P^\mu+\frac{f_{3}(q^{2})}{M_{\mathcal B_i}M_{\mathcal B_f}}\bar P^\alpha \bar P'^\mu
 +f_4g^{\alpha \mu}
 \right] \nonumber\\&&u(\bar P,S_z)\bar u(\bar{P},S_z)g_{\mu}^{\xi}\}.
\end{eqnarray}

After solving the Eqs. (A5), (A6), A(7) and A(8),
$f_1$, $f_2$, $f_3$, $f_4$ can be expressed by $F_1$, $F_2$, $F_3$ and $F_4$ which can be numerically evaluated through Eqs. (A1), (A2), A(3) and A(4). The  polarization sum formula for a particle with $S=3/2$ is

\begin{eqnarray}\label{s22}
\sum_{S_z}{u}_{\xi}(\bar P,S_z)\bar{u}_{\alpha}(\bar{P},S_z)=-(\bar{P}\!\!\!\!\!\slash+M_0)[T_{\xi\alpha}(\bar{P})-\frac{1}{3}\gamma^{\rho}T_{\rho\xi}(\bar{P})T_{\alpha\sigma}(\bar{P})\gamma^{\sigma}],
\end{eqnarray}
with
\begin{eqnarray}\label{s23}
T_{\xi\alpha}(\bar{P})=g_{\xi\alpha}-\frac{\bar{P}_{\xi}\bar{P}_{\alpha}}{M^2_0}.
\end{eqnarray}

\section{Semi-leptonic decay of  $\mathcal B_i(\frac{1}{2}^+)\to\mathcal B_f (\frac{3}{2}^+)  l\bar\nu_l$ }

The helicity amplitudes are expressed in terms of the form factors
for $\mathcal B_i(\frac{1}{2}^+)\to\mathcal B_f (\frac{3}{2}^+) $ \cite{Korner:1991ph,Bialas:1992ny}
\begin{eqnarray}
  \label{eq:haad}
  H^{V,A}_{1/2,\, 0}&=&\mp\frac1{\sqrt{q^2}}\frac{2}{\sqrt3}
\sqrt{M_{\mathcal B_i}M_{\mathcal B_f}(w\mp
  1)}[(M_{\mathcal B_i}w-M_{\mathcal B_f}){\cal N}^{V,A}_4(w)\cr
&&\mp (M_{\mathcal B_i}\mp M_{\mathcal B_f})(w\pm1){\cal
N}^{V,A}_1(w) +M_{\mathcal B_f}(w^2-1){\cal N}^{V,A}_2(w)\cr
&&+M_{\mathcal B_i}(w^2-1){\cal N}^{V,A}_3(w)],\cr
 H^{V,A}_{1/2,\, 1}&=&\sqrt{\frac23}\sqrt{M_{\mathcal B_i}M_{\mathcal B_f}(w\mp
  1)}[{\cal N}^{V,A}_4(w)-2(w\pm 1){\cal N}^{V,A}_1(w)],\cr
H^{V,A}_{3/2,\, 1}&=&\mp\sqrt{2M_{\mathcal B_i}M_{\mathcal
B_f}(w\mp
  1)}{\cal N}^{V,A}_4(w),
\end{eqnarray}
where again the upper (lower)  sign corresponds  to $V(A)$ , ${\cal
  N}^V_i\equiv g_i$, ${\cal N}^A_i\equiv f_i$ ($i=1,2,3,4$) and the
  $q^2$ is the lepton pair invariant mass.
The remaining helicity amplitudes  can be obtained using the
relation
$$H^{V,A}_{-\lambda',\,-\lambda_W}=\mp H^{V,A}_{\lambda',\, \lambda_W}.$$

Partial differential decay rates can be represented in the
following form
\begin{eqnarray}
  \label{eq:darlt}
 \frac{d\Gamma_T}{dw}&=&\frac{G_F^2}{(2\pi)^3} |V_{Q_1Q_2}|^2\frac{q^2
   M_{\mathcal B_f}^2\sqrt{w^2-1}}{12M_{\mathcal B_i}} [|H_{1/2,\, 1}|^2+
   |H_{-1/2,\, -1}|^2+|H_{3/2,\, 1}|^2+
   |H_{-3/2,\, -1}|^2],\cr
\frac{d\Gamma_L}{dw}&=&\frac{G_F^2}{(2\pi)^3} |V_{Q_1Q_2}|^2\frac{q^2
   M_{\mathcal B_f}^2\sqrt{w^2-1}}{12M_{\mathcal B_i}} [|H_{1/2,\, 0}|^2+
   |H_{-1/2,\, 0}|^2],
\end{eqnarray}
where $p_c=M_{\mathcal B_f}\sqrt{w^2-1}$ is the momentum of
$\mathcal B_f$ in the reset frame of $\mathcal B_i$.

The differential decay width of the $\mathcal B_i(\frac{1}{2}^+)\to\mathcal B_f (\frac{3}{2}^+)
l\bar\nu_l$   can be written as
\begin{eqnarray}
  \label{eq:darlt}
 \frac{d\Gamma}{dw}&=&\frac{d\Gamma_T}{dw}+\frac{d\Gamma_L}{dw}.
\end{eqnarray}

Integrating over the parameter $\omega$, we can obtain the total
decay width
\begin{eqnarray}
  \label{eq:darlt}
\Gamma=\int_1^{\omega_{\rm max}}d\omega\frac{d\Gamma}{dw},
\end{eqnarray}
where $\omega=v\cdot v'$ and the upper bound of the integration
$\omega_{\rm max}=\frac{1}{2}(\frac{M_{\mathcal B_i}}{M_{\mathcal
B_f}}+\frac{M_{\mathcal B_f}}{M_{\mathcal B_i}})$ is the
maximal recoil.

The ratio of the longitudinal to transverse decay rates $R$ is
defined by
 \beq
 R=\frac{\Gamma_L}{\Gamma_T}=\frac{\int_1^{\omega_{\rm
     max}}d\omega~q^2~p_c\left[ |H_{\frac{1}{2},0}|^2+|H_{-\frac{1}{2},0}|^2
     \right]}{\int_1^{\omega_{\rm max}}d\omega~q^2~p_c
     \left[ |H_{1/2,\, 1}|^2+
   |H_{-1/2,\, -1}|^2+|H_{3/2,\, 1}|^2+
   |H_{-3/2,\, -1}|^2\right]}.
 \eeq

\end{document}